\documentclass[review,number,sort&compress]{elsarticle}

\makeatletter
\def\ps@pprintTitle{%
 \let\@oddhead\@empty
 \let\@evenhead\@empty
 \def\@oddfoot{\centerline{\thepage}}%
 \let\@evenfoot\@oddfoot}
\makeatother

\usepackage{amssymb}
\usepackage{amsmath} %Sabrina

\usepackage{color}
\usepackage{lineno}

\usepackage{booktabs}

\begin{document}

\begin{frontmatter}



\title{Uncertainty limits on solutions of inverse problems over multiple orders of magnitude using bootstrap methods: An astroparticle physics example}

\author[label1]{Sabrina Einecke\corref{cor1}}
\ead{sabrina.einecke@tu-dortmund.de}
\author[label2]{Katharina Proksch\corref{cor1}\fnref{label3}}
\ead{katharina.proksch@ruhr-uni-bochum.de}
\author[label2]{Nicolai Bissantz}
\author[label1]{Fabian Clevermann}
\author[label1]{Wolfgang Rhode}

\address[label1]{Institute of Physics, Technische Universit\"at Dortmund, D-44221 Dortmund, Germany}
\address[label2]{Institute of Statistics, Ruhr-Universit\"at Bochum, D-44780 Bochum, Germany}

\cortext[cor1]{Corresponding authors.}

\fntext[label3]{Now at: Institute for Mathematical Stochastics, Georg-August-Universit\"at G\"ottingen, D-37077 G\"ottingen, Germany}

\begin{abstract}
Astroparticle experiments such as IceCube or MAGIC require a deconvolution of their measured data with respect to the response function of the detector to provide the distributions of interest, e.\,g.\ energy spectra.
In this paper, appropriate uncertainty limits that also allow to draw conclusions on the geometric shape of the underlying distribution are determined using bootstrap methods, which are frequently applied in statistical applications.
Bootstrap is a collective term for resampling methods that can be employed to approximate unknown probability distributions or features thereof.
A clear advantage of bootstrap methods is their wide range of applicability. For instance, they yield reliable results, even if the usual normality assumption is violated.

The use, meaning and construction of uncertainty limits to any user-specific confidence level in the form of confidence intervals and levels are discussed.
The precise algorithms for the implementation of these methods, applicable for any deconvolution algorithm, are given. 
The proposed methods are applied to Monte Carlo simulations to show their feasibility and their precision in comparison to the statistical uncertainties calculated with the deconvolution software TRUEE.
\end{abstract}

\begin{keyword}
Uncertainty limits \sep Bootstrap \sep Unfolding \sep Inverse problem



\end{keyword}

\end{frontmatter}

%\linenumbers

\section{Introduction}
\label{sec:intro}

A problem is called an inverse problem, if it is impossible to observe a quantity of interest directly, and only indirect observations are available for any kind of inference about it.
The process of solving an inverse problem is referred to as unfolding or deconvolution.
These kinds of problems arise in many research areas of science, economics and engineering.
In astroparticle physics, such a sought-after but not directly measurable feature of an observed particle is, e.\,g., the particle's energy.
In this case, the corresponding energy distribution is the quantity of main interest, it extends over multiple orders of magnitude both in energy and particle numbers.
These energy distributions -- also referred to as energy spectra -- are predicted by several theories, baring different insights into the universe.
In order to compare a measurement with these theories and to prove a certain theory, it is necessary to rely on uncertainty limits to distinguish between random fluctuations and substantial differences between a theory and the underlying spectra.
A frequently used method in statistical applications to construct these uncertainty limits is the bootstrap method, which is based on repetitive random sampling from a measurement to draw conclusions on features of its probability distribution.
It allows the calculation of pointwise confidence intervals as well as of confidence bands for any confidence level and is applicable for any unfolding algorithm.
Pointwise confidence intervals allow pointwise statements regarding the underlying spectrum, while confidence bands allow also for the validation of global statements regarding, e.\,g., the overall  geometric shape of the underlying spectrum.

In Section~\ref{sec:Ill posed inverse problems}, an overview of inverse problems and their solution is given, as well as an introduction to the unfolding software TRUEE (Time-dependent Regularized Unfolding for Economics and Engineering problems~\cite{Milke2012}).
In Section~\ref{sec:Confidence bars and confidence bands}, the differences between confidence intervals and confidence bands will be discussed.
The description of constructing uncertainty limits in terms of confidence intervals and bands on solutions of inverse problems using bootstrap methods will be described in Section~\ref{sec:methods}.
In Section~\ref{sec:appl}, a toy Monte Carlo simulation of a typical astroparticle example will be unfolded with the software TRUEE and the proposed bootstrap method will be applied.
Section~\ref{sec:disc} provides a discussion of the obtained results and conclusions.
\section{Inverse problems and unfolding}
\label{sec:Ill posed inverse problems}

A mathematical model that describes the folding process of the function of interest with a certain response function is given by the Fredholm integral equation~\citep{fredholm1903}

  \begin{equation}\label{foldingMath}
    g(y)=\int A(y,E)f(E)\,\text{d}E\ ,
  \end{equation}

where $f$ denotes the function to be recovered (e.\,g., the energy spectrum of particles).
The kernel $A$ of the integral equation corresponds to the response function of the measurement process (e.\,g.\ the detector) and $g$ is the function that can be observed.
A more detailed discussion of the model, its implications and origin can be found in~\cite{Milke2012}.

A typical data set only consists of a finite number of measurements, e.\,g.\ $N$ particles with their folded energies. 
Each data point contains random noise, which is passed on to the unfolded data. 
This process of unfolding is very sensitive to noise in the data, which is why this sort of problem is called an ill-posed problem~\citep{Blobel:2002pu}. 

The statistical model corresponding to the mathematical, idealized model (\ref{foldingMath}) is given by

  \begin{equation}\label{eq:foldingStat}
    Y_j=\int A(y_j,E)f(E)\,\text{d}E+\varepsilon_j, \quad j=1,\ldots,N.
  \end{equation}

The quantities $\varepsilon_1,\ldots,\varepsilon_N$ denote centered random uncertainties, i.\,e.\ uncertainties with  zero mean: $\mathbb{E}(\varepsilon_j)=0$ for $j=1,\ldots,N$.
The measurements $Y_1,\ldots,Y_N$ contain only indirect information about the function $f$ of interest. 
The problem differs from other problems where it is assumed that the function $f$ is directly and empirically accessible.
The difficulty in this context is that only data of the transformed function 

\begin{equation}\label{eq:foldingDef}
	g:=\int A(\cdot,E)f(E)\,\text{d}E
\end{equation}

are accessible.
The data need to be unfolded first.
In this context, unfolding means to recover the function~$f$ in the integral equation~(\ref{foldingMath})  as precisely as possible, if only discrete and noisy data as in ~(\ref{eq:foldingStat}) are available.
Unfolding will only give an approximation of the true underlying distribution. 
This estimation of $f$ will be denoted by $\hat f$ to distinguish between the true and the reconstructed spectrum.
To recover a particle spectrum, several algorithms are available, e.\,g.\ within the RooUnfold framework~\cite{roounfold}, or based on singular value decomposition as proposed in~\cite{svd}. 
All these algorithms are only capable to use one measured observable during the unfolding process. 
To study the proposed bootstrap algorithm for the calculation of uncertainty limits, the unfolding software TRUEE (Time-dependent Regularized Unfolding for Economics and Engineering problems~\cite{Milke2012}) will be used, providing stable results and reliable uncertainty estimations. 
TRUEE is able to use up to three measured observables during the unfolding fit, increasing the informative content and thus optimizing the estimation of the function to be recoverd.
\section{Confidence bars and confidence bands}
\label{sec:Confidence bars and confidence bands}

Methods for the recovery of a signal from (indirect) data, such as the recovery of the energy spectrum, are in general not sufficient for drawing quantitative conclusions on the (astrophysical) problem under consideration.
For example, in order to decide whether the reconstructed energy spectrum approaches zero with a certain slope towards high energies, an estimate on the uncertainty of the reconstruction of the spectra is needed. 
Otherwise, limits on the range of values for the slope consistent with the data cannot be provided. 
In this section, the basic concepts of both pointwise and uniform confidence intervals on the recovered signal are introduced, providing the required confidence intervals and bands, respectively.
Note that the specification of a confidence level $\alpha$ in mathematical notation corresponds to some multiple $k$ of $\sigma$ in physical notation.  
The precise connection is given via the equation $\alpha=2\,(1-\Phi(k))$, where $\Phi$ is the cumulative distribution function of the standard normal distribution. 
Accordingly, a level of $1\,\sigma$ corresponds to $\alpha=0.32$ and $\alpha=0.05$ corresponds to $1.96\,\sigma$.

\subsection{Basic statistical concept: Estimates of uncertainty}
\label{sub:basic statistical concept}

A confidence interval is always associated with a predefined probability, i.\,e.\ a small number $\alpha\in(0,1)$, which is needed to fix the level of the confidence interval. 
$(1-\alpha)$ is the (minimum) probability that at some fixed energy $E$ the true value $f(E)$ is actually contained in the confidence interval. 
With a probability of at most~$\alpha$, the confidence intervals are chosen either too small or at the wrong location and do not contain the true value $f(E)$.
Accepting such a residual risk is always necessary because of the stochastic nature of the data. 
Very small values of $\alpha$ yield a very small level of uncertainty, but at the cost of only a low precision, that is, large confidence intervals, which are often useless for any kind of inference. 
Acceptance of a larger level of residual risk (e.\,g.\ $\alpha=0.1\ldots 0.2$) yields small confidence intervals, resulting in a trade-off between information and certainty. 
In statistical applications, typical values of $\alpha$ are $0.01, \,0.05$ and $0.1$.

\subsection{Pointwise confidence intervals}
\label{sub:confidence bars}

If an energy spectrum $f(E)$ is reconstructed at energy intervals $E_1,\,E_2,\,\ldots,E_n$, a confidence interval can be constructed for each of these energy intervals $E_j$ separately. 
For each of the $n$ energy intervals where the spectrum is estimated, there is a chance $\lesssim\alpha$  that the true value of the spectrum is not covered by the respective confidence interval. 
Confidence intervals are in general (under some additional assumptions) expected to underestimate the true uncertainty for about $\alpha\cdot n$ points of the spectrum. 
The construction of $n$~different, separately constructed confidence intervals for $f(E_1),\ldots,f(E_n)$ can only be used to draw conclusions at the points $E_1,\ldots,E_n$, respectively. 
The probability, that the true spectrum is outside  of at least some of the pointwise confidence intervals in the considered range, is substantially larger than $\alpha$, it will approach $100\%$ with $n\to\infty$. 
The determination of the shape of the spectrum, e.\,g.\ the slope of the spectrum, is problematic.

\subsection{Uniform confidence bands}
\label{sub:confidenceBands}

It is possible to construct uniform confidence bands (upper and lower limiting spectra) that contain the complete reconstructed spectrum with probability $(1-\alpha)$. 
For instance, the probability, that a range of spectral slopes, which is determined based on the minimal and maximal slope consistent with the confidence band, does not include the true value, is only~$\alpha$.
The risk is much larger if the estimation is based on pointwise confidence intervals with a risk of failure of $\alpha$ for each energy where the spectrum was reconstructed.

In the literature, several standard approaches are available that can be used to construct uniform confidence bands. 
They are based on limit theorems for the maximum deviation of the reconstructed spectrum $\hat f(E)$ and the true spectrum $f(E)$ within an interval $[E_{\rm min},E_{\rm max}]$, i.\,e.\ a limit theorem that gives the limiting distribution of the properly scaled quantity

\begin{equation}
     Z:= \max_{E\in[E_{\rm min},E_{\rm max}]}\bigl|f(E)-\hat f(E)\bigr|,
     \label{eq:z}
\end{equation}

where the limit is taken as the number $N$ of observations used to obtain $\hat f$ (see model~(\ref{eq:foldingStat})) tends to infinity.
Theoretical results like this were first obtained by~\cite{smirnov1950} for histogram estimates and by~\cite{bicros1973a} for more general kernel density estimators. 
This latter method was further investigated and transferred to regression problems by~\cite{johnston1982} and~\cite{eubspe1993} among many others. 
None of those methods consider inverse problems, instead, all assume that the quantity of interest is directly observable. 
The first method of this type for inverse problems can be found in~\cite{bisdumholmun2007} in the context of density deconvolution and in~\cite{birbishol2010} for inverse regression problems. 
By means of the limit distribution of $Z$ (Eq. \eqref{eq:z}), it is possible to determine the width of the uniform confidence band $C(\alpha)$ in dependence of certain characteristics of the data such that the risk that the confidence band with center $\hat f(E)$ does not cover the true spectrum $f(E)$ is at most (approximately) $\alpha$:

\begin{equation} 
	\begin{split}
		P\left(\hat f(E) - C(\alpha)\leq f(E)\leq C(\alpha)+\hat f(E)\right)\geq (1-\alpha)\\ 
		\forall \; E\in[E_{\rm min},E_{\rm max}] \ . 
	\end{split}
\end{equation}

\subsection{Feasibility and alternatives}
\label{sub:practicality}

Unfortunately, results and methods for uniform confidence bands in the case of inverse problems are only available for the estimation of the density (i.\,e.\ frequency distribution) of some quantity of interest which is only observable with noise, and for the case of regression if the variance of the observation uncertainty does not depend on the independent variable in the regression. 
In the context of density estimation, the standardization in the limit theorem includes the original density $g$ (see model~(\ref{eq:foldingDef})). 
The reason is that in areas where the density of interest is large, more certain statements can be made because more data are available. 
In areas of small probability density, the certainty is small, resulting in wider bands. 
Methods to determine pointwise confidence intervals or confidence bands can only be applied if the density $g$ in model~(\ref{eq:foldingDef}) can be estimated reasonably well. 

The approximation of a distribution of some quantity $Z$ (Eq.\ (\ref{eq:z})) by its limit for moderate sample sizes can lead to a very poor performance in the results for moderate (practically available) sample sizes. 
In such a case, more reliable methods are of interest.
A general approach are bootstrap methods which often have the same desirable asymptotic properties as the methods based on limit distributions, but provide better finite sample performance, i.\,e.\ are more reliable when applied to real data. 
Bootstrap is a collective term for resampling methods that can be used to estimate distributions of statistical quantities or certain characteristics thereof. 
Bootstrap techniques are based on the idea to use a given data sample of size $N$ to create many new data sets comparable to the original one and, by performing the same analysis to these bootstrap datasets as to the real data, understand the statistical uncertainties and characteristics of the method of reconstruction. 

In addition to the above introduced methods, a third method will be used for the reconstruction of the density, which provides a compromise between pointwise confidence intervals and uniform confidence bands. 
For this method, pointwise confidence intervals are constructed with the Bonferroni method such that the chance, that the true spectrum violates the uncertainty limit in at least one point does not exceed $\alpha$, i.\,e.\ a discrete version of the uniform confidence bands. 
Such Bonferroni-corrected bands were already discussed in a non-parametric regression problem with a directly observable quantity of interest in~\cite{eubspe1993}, where it was also demonstrated that these bands tend to be larger than uniform confidence bands.
This is only critical for large numbers of values of the independent variable (i.\,e.\ $E_1,\ldots,E_n$ with $n\gtrsim 200$), where the spectrum and the associated pointwise confidence intervals are considered. 
The implementation of the method makes use of pointwise confidence intervals at level $\alpha'=\alpha/n$. 
The overall simultaneous level of failure of the confidence intervals is bounded from above by $\mathbf\alpha$, i.\,e.\ the constructed Bonferroni confidence bands are conservative. 
Unfortunately, for increasing number of points $E_1,\ldots,E_n$, these Bonferroni confidence bands are too wide by an increasing amount. 

\section{Method}
\label{sec:methods}

In the following, the bootstrap algorithm for the determination of the confidence intervals and bands from a sample of size $N$, i.\,e.\ the total number $N$ of measured particles, is described. 
The total number of bins, i.\,e.\ the number of points at which the energy spectrum is reconstructed, is denoted by $n$.
The index $i$ denotes the index of a data point, $E_i$ means the $i$-th value of interest, respectively, the $i$-th bin.
The bootstrap algorithm comprises $M$ iterations in total with $j$ denoting the number of the current iteration, i.\,e.\ $j=1,\ldots,M$. 
Note, that $M$ can be selected completely independent of $N$ and $n$. 

For all types of uncertainty limits discussed in this paper, the steps 0, 1 and 2 described below  are always the same, only step 3 is different for each method.

\noindent\textbf{Step 0:} Measurement of a data sample of size $N$.

\noindent\textbf{Step 1:} Determination of the reconstructed function $\hat f(E)$ by unfolding these $N$ data points at all points $E_1,\ldots,E_n$.

\noindent\textbf{Step 2:} Repetition of the following three steps $\bigl((i)-(iii)\bigr)$ a total number of $M$ times. The number $M$ should be large, i\,.e\,. $M\geq 500.$ 

\textbf{(i)} Obtaining $N$ data points by resampling with replacement from the measured data sample of Step 0. (The number $N$ in this step is equal to the number  $N$ of Step 0.)

\textbf{(ii)} Unfolding of this data sample to determine the function $\hat f_j^*(E)$ as in Step 1. 

\textbf{(iii)} Calculation of the deviations $S^i_j$ between the reconstructed function $\hat f(E)$ from Step 1 and the function $\hat f_j^*(E)$ for each bin $E_i$, i.\,e.\

  \begin{equation}
    S^i_j:= \left|\hat f(E_i)-\hat f_j^*(E_i)\right|
  \end{equation}  

and the largest deviation 

  \begin{equation}
    S_j:= \max_{i=1,\ldots,n}S^i_j\ .
  \end{equation}  

\noindent\textbf{Step 3:} Estimation of the confidence intervals and bands from the distributions $S^i_j$ and $S_j$ for level $\alpha$ as described below for the different methods.
  
\noindent\textbf{Pointwise confidence intervals:} Determination of the $(1-\alpha)$-quantile $q(i)_{(1-\alpha)}$ from the sample $S^i_1,\ldots,S^i_M$ of deviations from Steps 2 for each $i$ by rearranging the distances $S^i_j$ according to their value: $S^i_{(1)}\leq S^i_{(2)}\leq\ldots\leq S^i_{(M)}$.        
The quantity

  \begin{equation}
    q(i)_{(1-\alpha)}=
    \begin{cases}
      S^i_{(M\cdot (1-\alpha))} & \textnormal{if } M\cdot (1-\alpha)\in\mathbb{N}\\
      S^i_{(\left\lfloor M\cdot (1-\alpha) +1\right\rfloor)}& \textnormal{if } M\cdot (1-\alpha)\notin\mathbb{N}
    \end{cases}
    .
  \end{equation}

is a commonly used, natural estimate for the $(1-\alpha)$-quantile.
The pointwise confidence interval is then given by $\hat f(E_i)\pm q(i)_{(1-\alpha)}.$
     
\noindent\textbf{Uniform confidence band:} Determination of the $(1-\alpha)$-quantile $q^u_{(1-\alpha)}$ of the largest deviations only, i.\,e.\, the distribution of $S_j$, by replacing $ S^i_{(M\cdot (1-\alpha))}$ and $S^i_{(\left\lfloor M\cdot (1-\alpha)\right\rfloor}$ by $ S_{(M\cdot (1-\alpha))}$ and $S_{(\left\lfloor M\cdot (1-\alpha)\right\rfloor}$, respectively.  
The uniform confidence band is then given by $\hat f(E_i)\pm q^u_{(1-\alpha)}.$
    
\noindent\textbf{Bonferroni-corrected confidence band:} Determination of the $(1-\frac{\alpha}{n})$-quantile $q(i)_{(1-\frac{\alpha}{n})}$ of the distribution of $S^i_j$ in a similar way as described for $q(i)_{(1-\alpha)}$. 
The Bonferroni-corrected confidence band for level $\alpha$ is then given by $\hat f(E_i)\pm q(i)_{(1-\frac{\alpha}{n})}.$ 

\section{Application}
\label{sec:appl}

To illustrate its feasibility, the proposed method is applied to toy Monte Carlo simulations of a typical astroparticle example.
One Monte Carlo set is generated to determine the kernel for the unfolding, conducted with the software TRUEE.
Another simulated Monte Carlo set is used as pseudo data, i.\,e.\ this set is treated the same as real data, allowing a comparison between the result of the unfolding and the truth.
For a comparison of the obtained results with the well tested uncertainties of TRUEE and additional uncertainty estimations, the proposed method is slightly modified.
In addition, multiple pseudo data sets are generated independently for further test purposes.

\subsection{Monte Carlo simulation}
\label{sec:methods:montecarlo}

\begin{figure}[!h]
	\centering
	\includegraphics[width=0.9\textwidth]{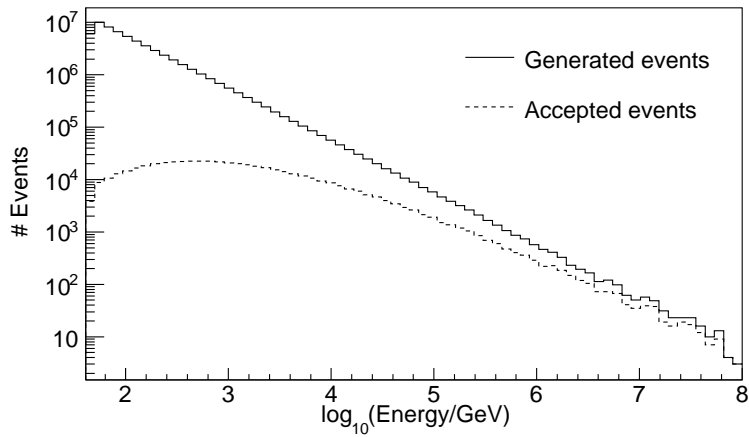}
	\caption{Energy distribution of the $60\,000\,000$ generated events (solid) and of the $494\,272$ accepted events (dashed).}
	\label{fig:GeneratedEvents}
\end{figure}

\begin{figure}[!h]
	\centering
	\includegraphics[width=.9\textwidth]{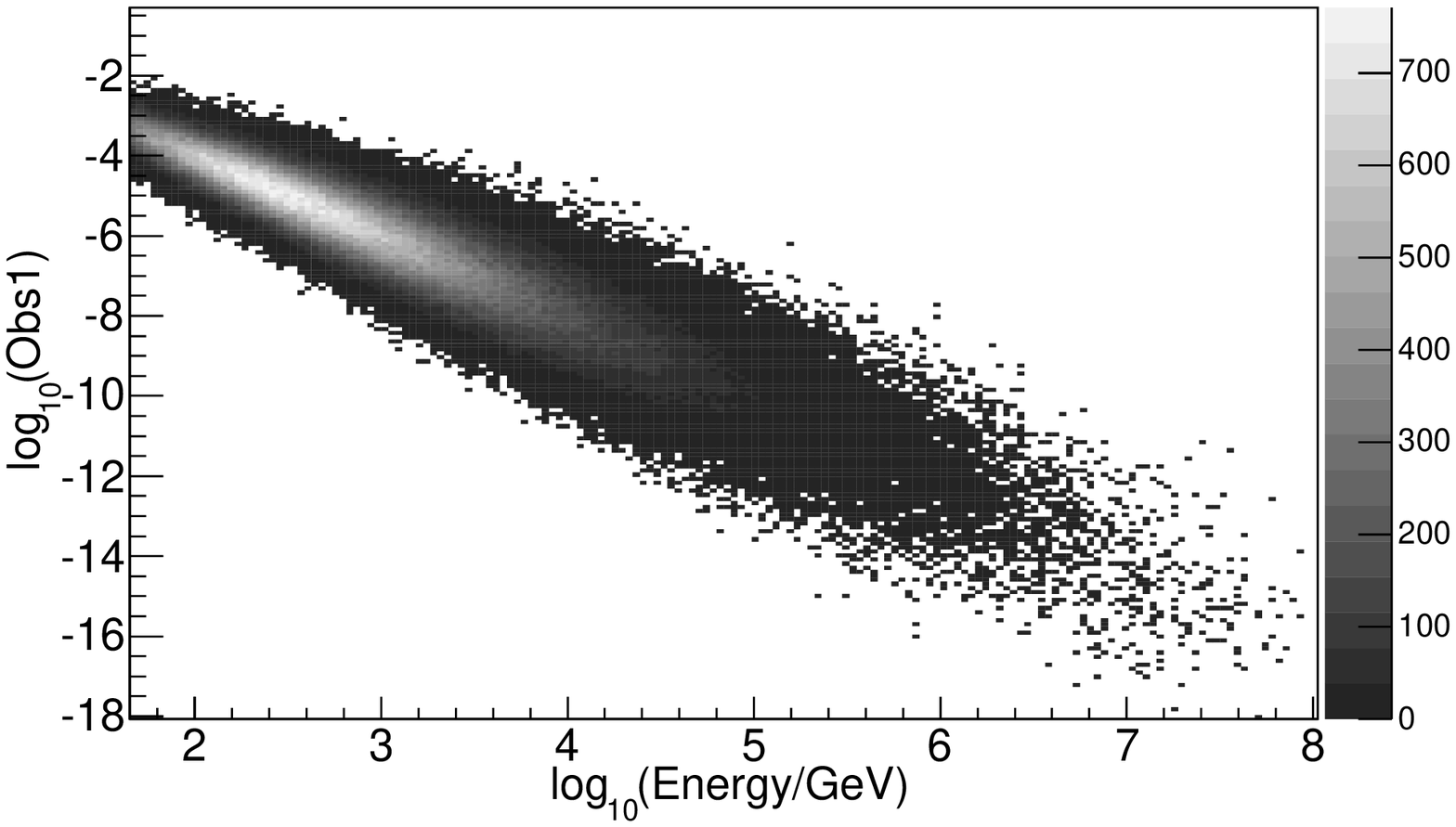}
	\includegraphics[width=.9\textwidth]{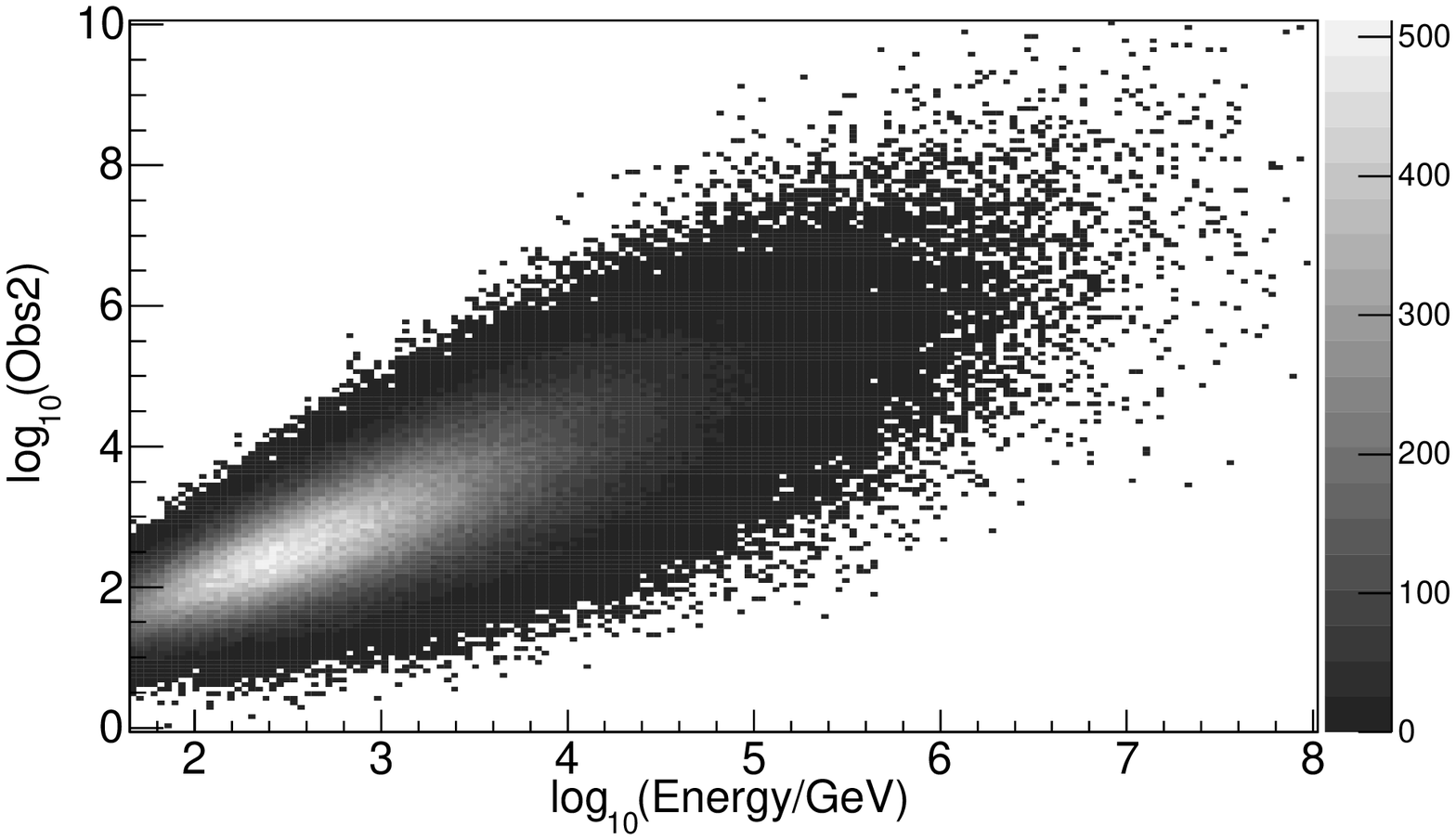}
	\caption{Dependence between the energy and the generated observables $Obs1$ (top) and $Obs2$ (bottom), respectively.}
	\label{fig:Obs1}
\end{figure}

The reconstruction of particle distributions over multiple orders of magnitude, such as the energy spectra of neutrinos or gamma rays, are typical problems in astroparticle physics.
The generated toy Monte Carlo includes an energy spectrum following a power law of $E^{-2}$ in the high-energy regime from $45\,$GeV to $100\,$PeV.
The simulation also includes an acceptance simulation, as most devices (e.\,g.\ the neutrino detector IceCube or the gamma ray telescope MAGIC) cannot detect every signal.
The used acceptance function $a(E)$ is

\begin{equation}
	a(E)=\left(1-\exp\left(\frac{- \log_{10}(\text{E})}{2}\right)\right)^{13}
\end{equation}
and its influence is shown in Figure~\ref{fig:GeneratedEvents}.
The distributions of simulated observables to be used in the unfolding procedure are similar as in real experiments, compare Figure~\ref{fig:Obs1} to observable distributions in \citep{Milke2012}.

As pseudo data $6\,000\,000$ events are generated resulting in $N = N_\text{pseudo} \approx 50\,000$ accepted events. 
For the determination of the kernel within the unfolding procedure, $60\,000\,000$ Monte Carlo events were generated, which resulted in $N_\text{MC} \approx 500\,000$ events.

\subsection{Unfolding}
\label{subsec:unfolding}

\begin{figure}
	\centering
	\includegraphics[trim=0cm 0cm 1cm 0cm, clip=true,width=0.99\textwidth]{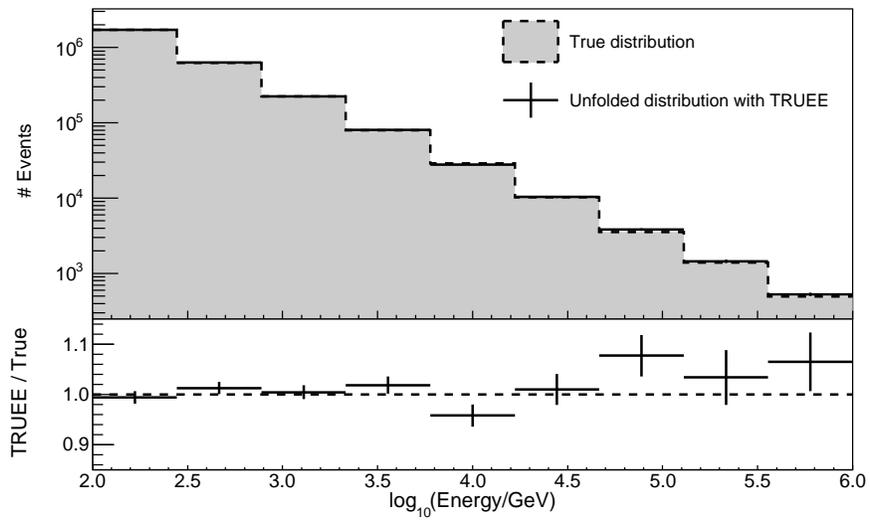}
	\caption[]{Comparison of the true energy distribution of events (dashed) and the unfolding result including the 68\% uncertainty, determined by the software TRUEE (solid). The ratio of these distributions is depicted in the lower figure and indicates an appropriate estimation of the spectrum and the corresponding uncertainties.}
	\label{fig:spectrum}
\end{figure}

The pseudo data set is unfolded in the energy range from $E_{\text{min}} = 100$\,GeV to $E_{\text{max}} = 1$\,PeV with the unfolding software TRUEE.
By taking advantage of the built-in test mode of TRUEE, a wide range of different settings are tested and compared, and the following settings are chosen: 
9 bins in energy, 12 knots (representing the knots of the interpolation of $f(E)$), 8 degrees of freedom (representing the regularization strength) and the observables $Obs1$ and $Obs2$ are used for the unfolding fit.
The resulting unfolded spectrum as well as the true distribution of events are depicted in Figure~\ref{fig:spectrum}.
Any unfolding is conducted with the same unfolding settings and the same Monte Carlo set to determine the kernel.
Only the pseudo data sets vary.

\subsection{Results}
\label{subsec:results}

For each test presented in this section, $M = 1\,000$ bootstrap iterations are performed with $j=1,\ldots,M$.
From these $M$ results the different uncertainty limits are determined.
A total number of $n=9$ bins in energy is considered, in which the Bonferroni correction is applicable.

As stated in Section~\ref{sub:confidenceBands}, the quantity for constructing uniform confidence bands (Eq.~(\ref{eq:z})) needs to be scaled properly.
In the considered astroparticle example, the spectrum $f(E)$ spans multiple orders of magnitude in event numbers, leading to large differences in event numbers from bin to bin.
By taking the absolute deviation between the reconstructed function $\hat f(E)$ and the function $\hat f^*_j(E)$ from the individual bootstrap iteration, the maximum deviation $S_j$ stems almost always from the bin with the maximum event numbers and the uniform confidence band will depend largely on this bin.
Transferring this band to the remaining bins, the estimated uncertainty will be large and no conclusions on the shape of the spectrum can be drawn.
In the case of using relative deviations instead, i.\,e.

\begin{equation}
	S^i_j:= \frac{\left|\hat f(E_i)-\hat f_j^*(E_i)\right|}{\hat f(E_i)}\ ,
\end{equation}
the deviations are in the same order of magnitude and thus, this scale is more suited.

\begin{figure}
	\centering
	\includegraphics[trim=0cm 0cm 1cm 0cm, clip=true,width=0.99\textwidth]{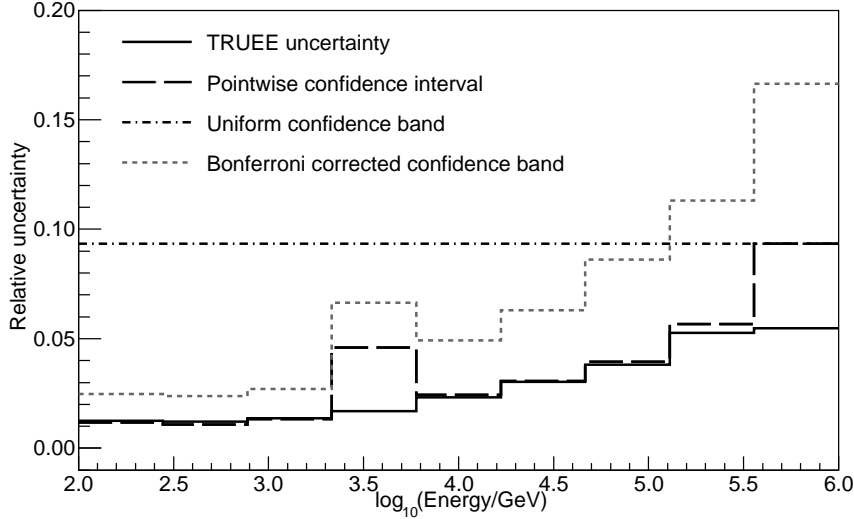}
	\caption[]{Relative 68\% confidence limits calculated by different approaches for resampled pseudo data sets. The pointwise confidence interval (dashed), the uniform (dash-dotted) and the Bonferroni corrected (dotted) confidence band are shown. The uncertainty of the unfolding software TRUEE is depicted as well (solid).}
	\label{fig:CR_deviationToF}
\end{figure}

The calculated relative 68\% confidence limits for $1\,000$ resampled data sets are depicted in Figure~\ref{fig:CR_deviationToF}.
As expected, the uniform and the Bonferroni corrected confidence band are larger as the pointwise confidence interval.
In the considered example, the Bonferroni corrected confidence band is more suited than the uniform confidence band, since it allows more precise conclusions, e.\,g.\ on the geometric shape of the spectrum of interest.
The value of the uniform confidence band is nearly the same as the pointwise confidence interval of the highest energy bin, which contains the fewest entries.
The largest relative deviations arise mostly in the bin with the fewest entries, the probability density is small and thus the uncertainty is large.
The effect of increasing uncertainty with decreasing event numbers is also visible for the TRUEE uncertainties.

\begin{figure}
	\centering
	\includegraphics[trim=0cm 0cm 1cm 0cm, clip=true,width=0.99\textwidth]{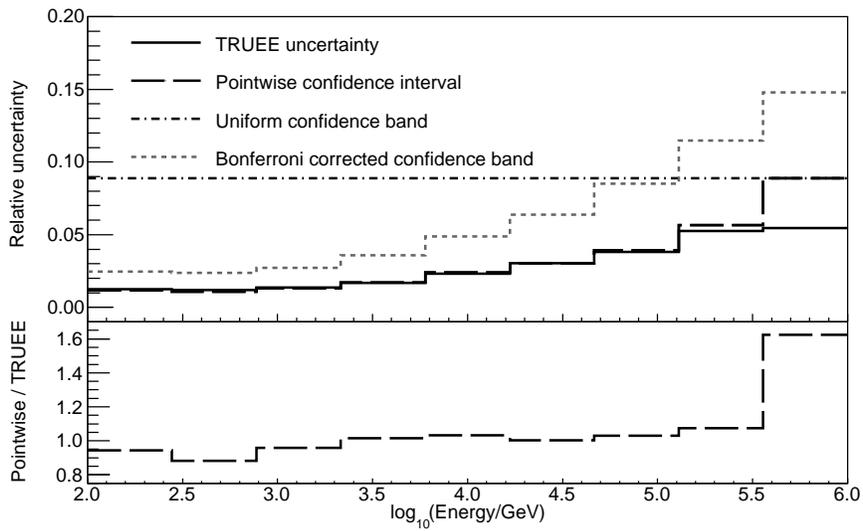}
	\caption[]{Relative 68\% confidence limits calculated by different approaches for resampled pseudo data sets. The pointwise confidence interval (dashed), the uniform (dash-dotted) and the Bonferroni corrected (dotted) confidence band are shown. The uncertainty of the unfolding software TRUEE is depicted as well (solid). In the lower figure the comparison between the pointwise confidence intervals and the uncertainties calculated by TRUEE is illustrated by their ratio. Values greater than $1$ indicate that the pointwise confidence interval is larger than the one of TRUEE.}
	\label{fig:CR_deviationToMedian}
\end{figure}

The standard bootstrap algorithm, described in Section~\ref{sec:methods}, estimates the standard deviation (a measure of dispersion) in combination with the bias (a measure of a systematic uncertainty).
As the software TRUEE calculates only the standard deviation, a slight modification of the bootstrap algorithm is performed for a qualitative comparison between the TRUEE uncertainties and the pointwise confidence intervals.
This modification concerns only the determination of $S^i_j$ in Step 2 (iii) as follows:

  \begin{equation}
    S^i_j:= \frac{\left|\tilde f(E_i)-\hat f_j^*(E_i)\right|}{\tilde f(E_i)}
  \end{equation}

with the medians $\tilde f(E_1)$, ..., $\tilde f(E_n)$ of the distributions $\hat f_j^*(E_1)$, ..., $\hat f_j^*(E_n)$.
The median is a more robust estimate of the center of the underlying distribution of the data.
Figure~\ref{fig:CR_deviationToMedian} represents the relative 68\% confidence limits for resampled data sets with the modified $S^i_j$.
In the bottom panel, the ratio between the pointwise confidence interval and the TRUEE uncertainty shows an agreement within about 10\% for almost every bin.
The bin with the fewest entries features a relatively large difference.

\begin{figure}
\centering
	\includegraphics[trim=0cm 0cm 1cm 0cm, clip=true,width=0.99\textwidth]{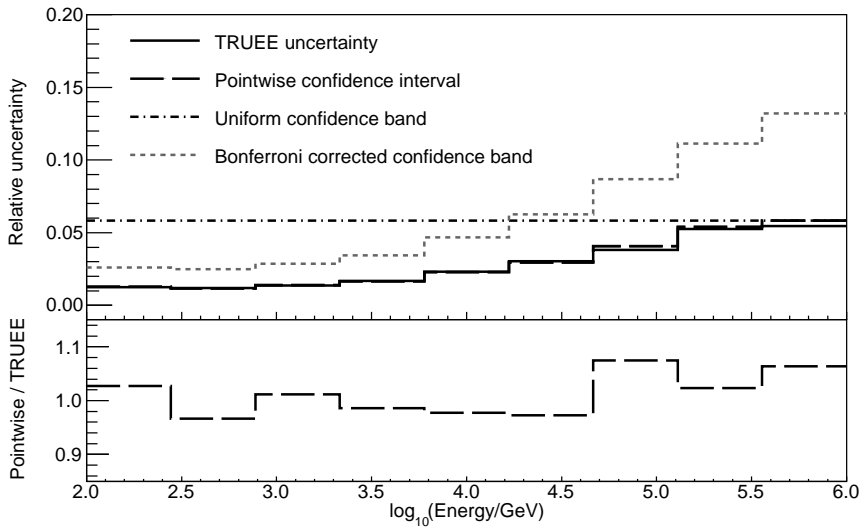}
	\caption[]{Relative 68\% confidence limits similar to Figure~\ref{fig:CR_deviationToMedian}. Here, $1\,000$ data sets are sampled from the true probability density function instead of the resampling from one single sample of the true probability density function.}
	\label{fig:1000pseudo_deviationToMedian}
\end{figure}

To investigate this large difference, the $1\,000$ sets are sampled from the true underlying  probability density function instead of resampling from one single sample of the probability density function and the bootstrap is performed again.
It should be noted that sampling from the true probability density function is only possible because toy Monte Carlo simulations are used.
The use of these data sets leads to an agreement within 8\% for each bin (see Figure~\ref{fig:1000pseudo_deviationToMedian}).
This proves the feasibility of the method in case of a sufficiently large probability density and demonstrates the constraints of the method for areas in which the probability density is very small.

\begin{figure}
\centering
	\includegraphics[trim=0cm 0cm 1cm 0cm, clip=true,width=0.99\textwidth]{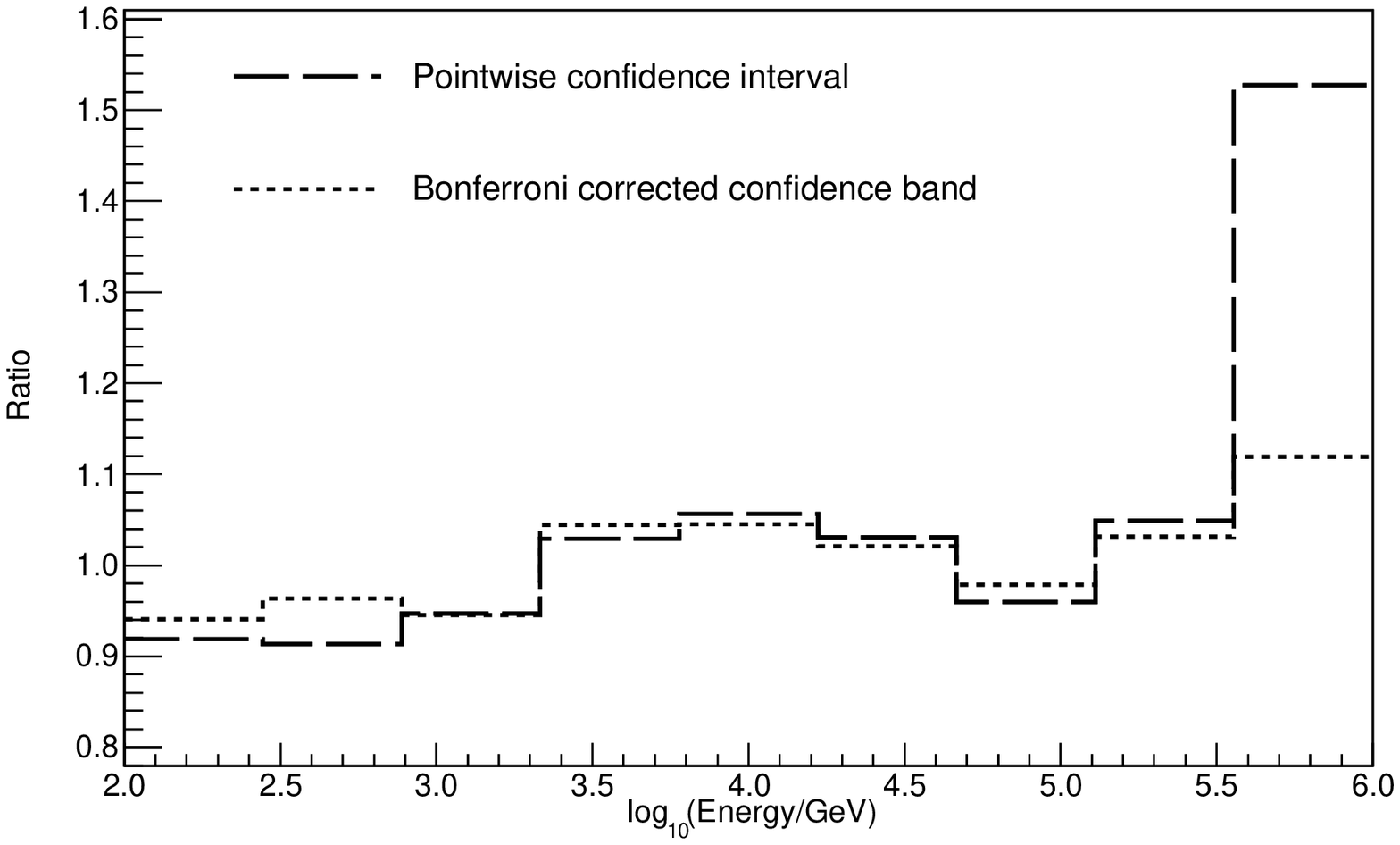}
	\caption[]{Comparison of uncertainties obtained with data sets sampled from the true probability density function and data sets resampled with replacement from one sample of the probability density function.}
	\label{fig:comparison_CR_1000pseudo}
\end{figure}

Figure~\ref{fig:comparison_CR_1000pseudo} illustrates the comparison between the uncertainties obtained with data sets sampled from the true probability density function (as in Figure~\ref{fig:1000pseudo_deviationToMedian}) and data sets resampled with replacement from one sample of the probability density function (as in Figure~\ref{fig:CR_deviationToMedian}).
The Bonferroni corrected confidence bands are consistent within 10\%.
The pointwise confidence intervals are compatible within 10\% as well, except for the bin with the smallest probability density.
Bonferroni corrected confidence bands turn out to be most robust against changes in the order of magnitude of the underlying probability density.

\section{Conclusions}
\label{sec:disc}

A bootstrap-based method for the construction of uncertainty limits on solutions of inverse problems has been presented, providing the possibility to determine both confidence intervals and bands to any user-specific confidence level.
This method is applicable for any unfolding algorithm and it might even improve the construction of uncertainty limits which are already implemented in an algorithm.
Standard methods to calculate uncertainty limits often rely heavily on assumptions regarding the (often normal or Poisson) distribution of the data.
The proposed bootstrap algorithm works independently from such strong restrictions.
Even if methods fail, because there these strong assumptions are violated, the less restrictive bootstrap-based methodology still works properly in many situations.
The uncertainty limits also include an estimate of the systematic uncertainty, describing the bias introduced by a regularization in the unfolding algorithm.

The method has been tested with a typical astroparticle physics example.
The use of relative instead of absolute deviations as an appropriate scale within the limit theorem for constructing uniform confidence bands has been proposed for problems extending over multiple orders of magnitude, yielding reasonable results for the calculation of confidence intervals as well as confidence bands.
It has been proven that the confidence intervals match the well checked uncertainties calculated by the unfolding software TRUEE.

Many applications where distributions are reconstructed can benefit from this bootstrap method.
This could be mass spectra in accelerator experiments, searching for new particles, as well as energy distributions used for the search of dark matter or the neutrinoless double beta decay.
It is often crucial to determine specific values or distributions with a particular confidence limit, e.\,g.\ for the detection of a new particle a confidence of 5$\,\sigma$ (corresponding to $\alpha\approx5.733\cdot10^{-7}$) is necessary.
The presented bootstrap method provides very useful tools, especially for physical applications, and extends the possibilities of existing unfolding algorithms and software.
\section*{Acknowledgment}

This research is supported by the German Research Foundation (DFG) through the Collaborative Research Center SFB 823, project C4, which is gratefully acknowledged.



\bibliographystyle{elsarticle-num} 
\bibliography{astrobib}

\end{document}